# Rapid MRI-Based Synthetic CT Simulations for Precise tFUS Targeting

Hengyu Gao[1], Shaodong Ding[2], Ziyang Liu[2], Jiefu Zhang[1], Bolun Li[1], Zhiwu An[5], Li Wang[2,3], Jing Jing[6], Tao Liu[2,3]*, Yubo Fan[1,2,3]* and Zhongtao Hu[1,3,4]*

1, School of Engineering Medicine, Beihang University, Beijing 100191, China.

2, School of Biological Science and Medical Engineering, Beihang University, Beijing 100191, China.

3, Beijing Advanced Innovation Center for Biomedical Engineering, Beijing 100191, China.

4, Shenzhen Institute of Beihang University, Shenzhen 518000, China.

5, State Key Laboratory of Acoustics and Marine Information, Institute of Acoustics, Chinese Academy ofSciences, Beijing 100162, China.

6, China National Clinical Research Center for Neurological Diseases, Beijing Tiantan Hospital, Capital Medical University, Beijing 100162, China.

Hengyu Gao and Shaodong Ding contributed equally to this work.

*Corresponding author: Zhongtao Hu (zhongtaohu@buaa.edu.cn), Tao Liu (tao.liu@buaa.edu.cn), and Yubo Fan (yubofan@buaa.edu.cn)

## 0. Abstract:

Accurate targeting is critical for the effectiveness of transcranial focused ultrasound (tFUS) neuromodulation. While CT provides accurate skull acoustic properties, its ionizing radiation and poor soft tissue contrast limit clinical applicability. In contrast, MRI offers superior neuroanatomical visualization without radiation exposure but lacks skull property mapping. This study proposes a novel, fully CT free simulation framework that integrates MRI-derived synthetic CT (sCT) with efficient modeling techniques for rapid and precise tFUS targeting. We trained a deep-learning model to generate sCT from T1-weighted MRI and integrated it with both full-wave (k-Wave) and accelerated simulation methods, hybrid angular spectrum (kWASM) and Rayleigh–Sommerfeld ASM (RSASM). Across five skull models, both full-wave and hybrid pipelines using sCT demonstrated sub-millimeter targeting deviation, focal shape consistency (FWHM ~3.3–3.8 mm), and <0.2 normalized pressure error compared to CT-based gold standard. Notably, the kW-ASM and RS-ASM pipelines reduced simulation time from ~3320 s to 187 s and 34 s respectively, achieving ~94% and ~90% time savings. These results confirm that MRI-derived sCT combined with innovative rapid simulation techniques enables fast, accurate, and radiation-free tFUS planning, supporting its feasibility for scalable clinical applications.

**Highlights:**

- Integrating MRI-derived synthetic CT with rapid acoustic simulation methods enables transcranial focused ultrasound planning without the need for CT imaging.

- Two hybrid simulation approachs (kW-ASM and RS-ASM) achieve accuracy comparable to conventional full-wave k-Wave simulations while significantly reducing computation time.

- The MRI-derived sCT approach provides precise targeting accuracy, validated against CT-based models, demonstrating its feasibility for clinical tFUS applications.

**Introduction**

Transcranial focused ultrasound (tFUS) is a rapidly advancing non-invasive technique used for neuromodulation and targeted deep brain stimulation. tFUS is a transformative technology with diverse applications in neuroscience and medicine. In neuromodulation, tFUS offers non-invasive treatments for essential tremor [1,2], Parkinson's disease[3–7], Alzheimer disease[8–13], and depression[14–16]. Beyond neuromodulation, tFUS facilitates targeted drug delivery by temporarily disrupting the blood-brain barrier to enable precise delivery of therapies for brain tumors[17–23], neurodegenerative diseases[24–26], and genetic or nanoparticle-based interventions[27–30]. High-intensity tFUS supports thermal ablation[31,32], while histotripsy provides a non-thermal, cavitation-based approach to mechanically disrupt tumor tissues[33,34]. Additionally, tFUS has enabled liquid biopsy applications, allowing the non-invasive detection of tumor or neurodegenerative biomarkers via enhanced release into the bloodstream[35–37]. The clinical effectiveness of tFUS hinges on the precision of ultrasound wave focusing, which requires accurate modeling of propagation through the skull—a heterogeneous structure with complex acoustic properties such as varying bone density and speed of sound [38–41,41,42]. Traditionally, computed tomography (CT) imaging has been the gold standard for extracting these properties due to its high-resolution depiction of bone density and structure [43,44]. However, CT's reliance on ionizing radiation, high costs, and limited suitability for repeated or longitudinal studies restrict its use[45–49], particularly for vulnerable populations such as healthy volunteers or patients requiring multiple sessions [50].

Magnetic resonance imaging (MRI) offers a radiation-free alternative with superior soft tissue contrast, making it ideal for neuroimaging and treatment planning. While conventional MRI does not directly provide skull acoustic properties, advancements in deep learning have enabled the synthesis CT (sCT) images from MR data. Various architectures, including 2D/3D convolutional neural networks (CNNs) and generative adversarial networks (GANs), have demonstrated the ability to generate sCT images with high fidelity to actual CT scans [51,52]. For example, Koh et al. developed a 3D conditional GAN model to generate sCT from T1-weighted MR images of 33 subjects [53], while Yaakub et al. introduced a toolbox based on a 3D residual U-Net trained on data from 110 subjects [54]. In 2019, Guo et al compared three methods for generating sCT images from MR data, illustrating that deep learning models, especially those utilizing UTE/ZTE MR sequences, can closely replicate CT accuracy for ultrasound therapy planning, reducing errors in acoustic energy estimation and focal precision[55]. Similarly, Su et al. utilized UTE sequences, employing a 2D CNN-based model that achieved less than 2°C error in acoustic and biothermal simulations[56]. More recently, Leung et al. evaluated the use of MRI-derived sCT for correcting skull-induced phase aberrations during tFUS [57], while demonstrated comparable accuracy between MR- and CT-based acoustic simulations by employing both classical and learned MR-to-sCT mappings [58]. Furthermore, sCT has also been applied successfully in MR-HIFU planning for bone metastases treatment [59]. In 2023, Liu et al. validated MRI-derived sCT for tFUS planning, showing its comparable performance to conventional CT in 10 independent test cases [60]. Collectively, these studies have demonstrated the ability of sCT to replicate the acoustic properties of CT, enabling its use in ultrasound propagation simulations. However, they rely heavily on traditional, computationally intensive simulation methods, which limit their feasibility for routine clinical use. Existing workflows emphasize accuracy but overlook the need for computational efficiency, posing challenges for widespread adoption in clinical applications. To address the high computational cost of traditional transcranial ultrasound simulations, rapid frameworks like the Angular Spectrum Method (ASM)[61–67] and Rayleigh-Sommerfeld diffraction integral[68–70] have been developed, offering significant improvements in efficiency while maintaining accuracy. Despite these advances, the integration of rapid simulation methods with MRI-derived sCT remains limited, leaving a gap in the development of efficient and practical workflows for routine clinical tFUS applications.

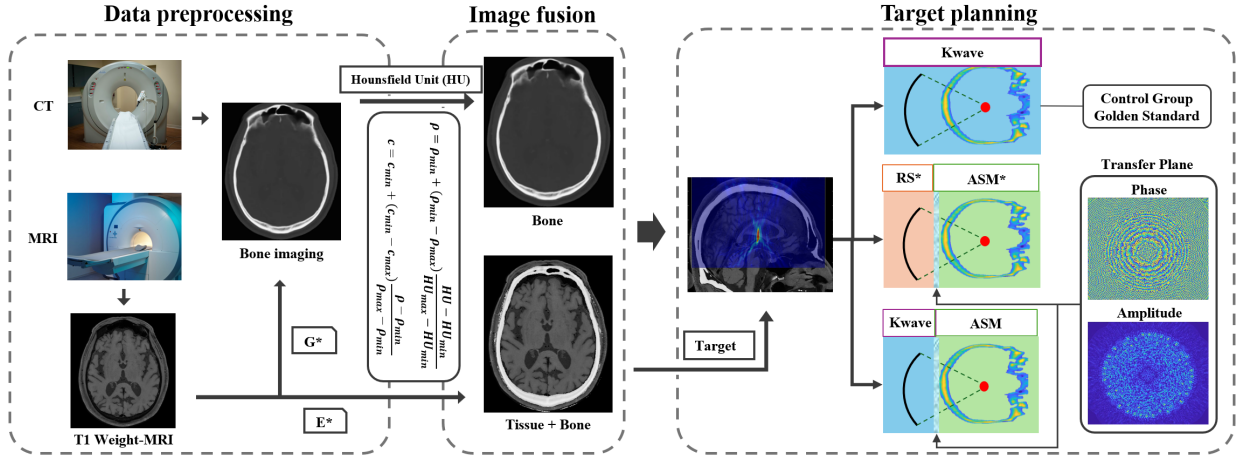

**Figure 1. Workflow Comparison of MRI-derived and CT-based tFUS Planning.** (1) Data Acquisition and Preprocessing, with CT providing HU values and MRI generating sCT is a radiation-free alternative; (2) Image Fusion, combining skull acoustic properties and tissue information for modeling; and (3) Target Planning, comparing k-wave (gold standard), RS-ASM (Rayleigh-Sommerfeld diffraction integral combined with the Angular Spectrum Method), and kW-ASM (k-wave combined with the Angular Spectrum Method) simulations in terms of accuracy and efficiency, with phase and amplitude distributions at the transfer plane.

In this study, we propose a CT-free, rapid tFUS simulation framework that combines MRI-derived sCT with two computationally efficient modeling methods: the k-Wave Angular Spectrum Method (kW-ASM) and the Rayleigh-Sommerfeld Angular Spectrum Method (RS-ASM). We evaluate this approach by comparing sCT-based simulations to gold-standard CT-based k-Wave models in terms of spatial accuracy, beam fidelity, and computation time. Our goal is to enable fast, accurate, and radiation-free ultrasound field simulation to support individualized and scalable tFUS treatment planning.

1. **Method**

**1.1 Generation of Synthesis CT from MRI**

We developed a deep learning pipeline to sCT images from paired T1-weighted (T1w) MRI data, achieving high consistency in bone Hounsfield Unit (HU) values compared to reference CT images. The pipeline includes data preprocessing, a U-Net-based network architecture, and training objectives.

**1.1.1 Data Preprocessing**

T1w images were preprocessed by removing the neck region, applying N4 bias field correction, and performing skull stripping using SynthStrip [71,72]. CT images underwent similar preprocessing, including OTSU thresholding to remove noise and registration to T1w images using ANTsPy[73]. Intensities were normalized (T1w via z-score, CT clamped to [-1024, 2000]). The final dataset included 22 paired T1w and CT volumes, resampled to 0.4×0.4×1.0 mm³ and resized to 576×576×192. Seventeen pairs were used for training, and five for testing.

**1.1.2 Network Architecture**

As shown in Figure 2, the CT synthesis network is an encoder-decoder U-Net [74] designed to generate CT image patches from T1w MRI patches. The encoder uses four 3D convolutional layers (stride = 2) to extract multi-scale features, followed by two vision transformer blocks [75] to enhance spatial feature aggregation and overcome the limitations of convolutional receptive fields. Multi-skip connections aid the decoder in reconstructing fine details, while subpixel upsampling [54] ensures smooth and realistic synthesized images.

**1.1.3 Training Objectives**

The network was trained to sCT images with HU values consistent with real CT (rCT) images, focusing on bone and soft tissue regions critical for tFUS. Loss functions included:

- **Global MAE** ($L_{MAE}^g$) to align all voxels with reference CT:

$$L_{MAE}^g = \frac{1}{H \times W \times D} \sum_{i=1}^{H} \sum_{j=1}^{W} \sum_{k=1}^{D} |o_{i,j,k} - y_{i,j,k}|. \qquad (1)$$

Where, $(H, W, D)$ is the shape of the image, $(i, j, k)$ is the spatial index of image, $o$ is the sCT image, and $y$ is the rCT image.

- **Bone MAE** ($L_{MAE}^{bone}$) focusing on bone voxels (HU: [100, 1500]):

$$L_{MAE}^{bone} = \frac{1}{V_{bone}} \sum_{i=1}^{H} \sum_{j=1}^{W} \sum_{k=1}^{D} \mathcal{M}_{bone} \times |o_{i,j,k} - y_{i,j,k}|. \qquad (2)$$

Here, $\mathcal{M}_{bone}$ is a binary mask identifying bone voxel.

- **Brain MAE** ($L_{MAE}^{brain}$) targeting soft tissue voxels (HU: [0, 100]):

$$L_{MAE}^{brain} = \frac{1}{V_{brain}} \sum_{i=1}^{H} \sum_{j=1}^{W} \sum_{k=1}^{D} \mathcal{M}_{brain} \times |o_{i,j,k} - y_{i,j,k}|, \qquad (3)$$

where $\mathcal{M}_{brain}$ is a binary mask identifying brain tissue voxels.

- **Global SSIM** ($L_{SSIM}^{g}$) to enhance structural similarity:

$$L_{SSIM}^{g} = 1 - \frac{(2\mu_o \mu_y + c_1)(2\sigma_{oy} + c_2)}{(\mu_o^2 + \mu_y^2 + c_1)(\sigma_o^2 + \sigma_y^2 + c_2)}, \qquad (4)$$

where $\mu_o$ and $\mu_y$ denote the mean values of the sCT and rCT images, respectively, while $\sigma_o, \sigma_y$, and $\sigma_{oy}$ represent the variance of the sCT, the variance of the rCT, and their covariance. Constants $c_1$ and $c_2$ are respectively set to 0.01 and 0.03 by default.

- **Brain SSIM** ($L_{SSIM}^{brain}$) to improve soft tissue structural fidelity:

$$L_{SSIM}^{brain} = 1 - \mathcal{M}_{brain} \times \frac{(2\mu_o \mu_y + c_1)(2\sigma_{oy} + c_2)}{(\mu_o^2 + \mu_y^2 + c_1)(\sigma_o^2 + \sigma_y^2 + c_2)}. \qquad (5)$$

The total loss function was defined as:

$$\begin{aligned} L = &\lambda_{MAE} L_{MAE}^{g} + \lambda_{SSIM} L_{SSIM}^{g} \\ &+ \lambda_{scale}(\lambda_{MAE} L_{MAE}^{bone} + \lambda_{MAE} L_{MAE}^{brain} + \lambda_{SSIM} L_{SSIM}^{brain}). \end{aligned} \qquad (6)$$

Here $\lambda_{MAE}, \lambda_{SSIM}$, and $\lambda_{scale}$ are hyperparameters and are set as 20, 5, and 0.5, respectively.

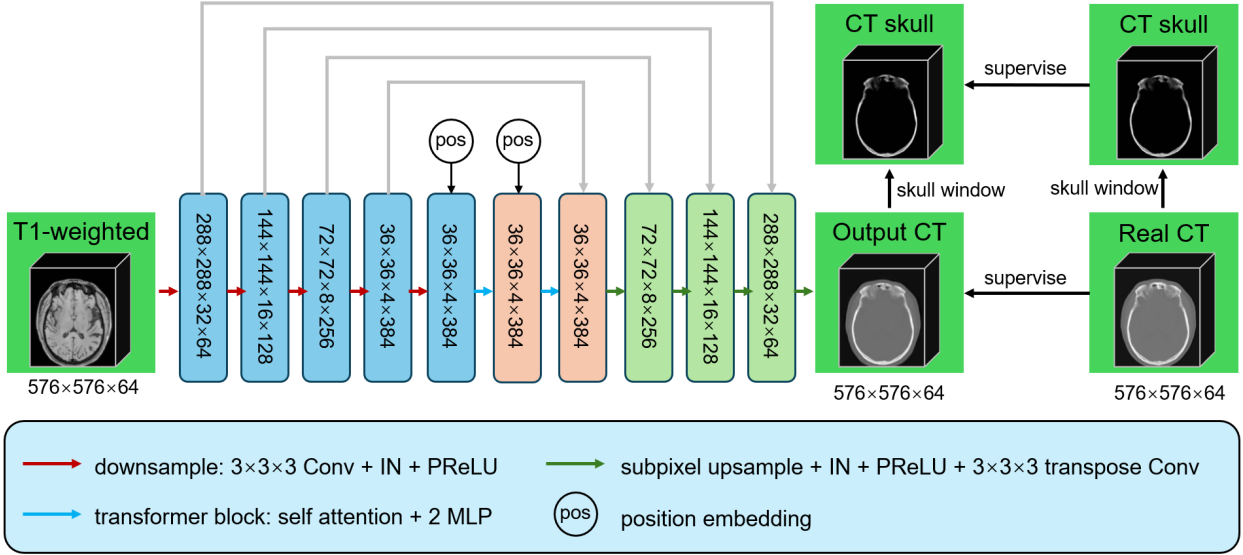

**Figure 2. Overview of the CT synthesis network, implemented as an encoder-decoder U-Net.** The network inputs a T1w image and outputs a sCT image. The encoder comprises four downsampling convolutional blocks, followed by two transformer blocks for enhanced spatial feature aggregation using position embeddings. The decoder reconstructs the CT image through four upsampling subpixel blocks, integrating feature maps from skip connections. 'Conv' indicates convolution, 'IN' instance normalization, and 'MLP' a multi-layer perceptron.

The model was implemented in PyTorch 2.0.1 and MONAI 1.3.0, trained on an NVIDIA 3090 GPU using the Adam optimizer with betas of 0.9 and 0.999, a learning rate of 3e-3, batch size of 2, and a cosine annealing scheduler. During training, we randomly cropped paired T1w and real CT images to 576×576×64 and trained the network for 2,000 epochs. For inference, we used the sliding window strategy in MONAI to synthesize a complete CT image.

### 2.2 MRI-Derived Acoustic Parameter Estimation

Simulations were performed using an open-source MATLAB toolbox, k-Wave, a pseudospectral method with k-space dispersion correction [76–79]. A graphics processing unit (Nvidia Tesla V100, Nvidia Corporation, Santa Clara, CA, USA) was used to accelerate the 3D simulations. The density ($\rho$) and sound speed ($c$) of the skull were converted from the Hounsfield units (HU) of the rCT or sCT images following the methods reported by Marsac et al.[41] as follows,

$$\rho = \rho_{\min} + (\rho_{\min} - \rho_{\max}) \frac{\text{HU} - \text{HU}_{\min}}{\text{HU}_{\max} - \text{HU}_{\min}}, \tag{7}$$

$$c = c_{\min} + (c_{\min} - c_{\max}) \frac{\rho - \rho_{\min}}{\rho_{\max} - \rho_{\min}}, \tag{8}$$

The density data of the skull ranges between $\rho_{min} = 1000$ kg/m³, and $\rho_{max} = 3200$ kg/m³, with an average density of $\rho_{mean} = 1525$ kg/m³, and the sound speed of skull ranges between $c_{min} = 1480$ m/s and $c_{max} = 4050$ m/s, with an average sound speed of $c_{mean} = 2542$ m/s, matching those reported in the previous literature [41]. The attenuation was obtained by the power law model as proposed by Aubry and Constans [80,81] as

$$Atten = \alpha_0 f^b \times \left(\frac{\rho - \rho_{min}}{\rho_{max} - \rho_{min}}\right)^\beta, \tag{9}$$

where $\alpha_0 = 8$ dB/cm/MHz, $b = 1.1$, $\beta = 0.5$, and $f$ is the central frequency. A numerical grid with an isotropic spatial resolution of 0.2 mm was generated. A numerical temporal step of $\Delta t = 66.7$ ns was used. The Courant-Friedrichs-Lewy number was 0.5 and the spatial sampling was approximately 15 grid points per wavelength in water at 500 kHz. These parameters were fixed for all simulations in this study. In contrast, the time-independent ASM and RS methods directly computed steady-state solutions, eliminating the need for iterative time-stepping. By standardizing the spatial resolution and sampling criteria across all methods, we ensured consistent accuracy while highlighting the computational efficiency of ASM and RS.

**2.3 Simulation methods**

This study employed three computational approaches to simulate acoustic wave propagation for tFUS targeting: (1) k-Wave simulation as the gold standard, (2) kW-ASM, and (3) RS-ASM. Each method was used to calculate the acoustic field, starting from a curved transducer surface and propagating through the skull to the target region (Figure 3). The following sections detail the methods and their governing equations.

**2.3.1 k-Wave Simulation method**

k-Wave is a full-wave simulation toolbox based on the k-space pseudo spectral method. It solves the first-order acoustic wave equations and accurately models sound wave propagation through heterogeneous media, such as the skull (Figure 3B). This method is considered the gold standard for transcranial ultrasound simulations due to its precision in capturing refraction, reflection, and absorption effects in complex anatomical regions[46,82–85]. The governing equations in k-Wave include Particle Velocity Equation, Density Equation and Pressure Equation as the following

$$\frac{\partial \mathbf{u}}{\partial t} = -\frac{1}{\rho}\nabla p, \tag{10}$$

$$\frac{\partial \rho_0}{\partial t} = -\rho_0 \nabla \cdot \mathbf{u}, \tag{11}$$

$$\frac{\partial p}{\partial t} = -c^2 \rho_0 \nabla \cdot \mathbf{u} - b \left(\frac{\partial p}{\partial t}\right)^{\frac{1}{2}}, \tag{12}$$

where, **u** is particle velocity vector (m/s). $p$ is the acoustic pressure (Pa). $\rho_0$ is the static density of the medium (kg/m³). $c$ is the speed of sound in the medium (m/s). $b$ is the absorption coefficient, related to the medium's attenuation properties. The pressure equation can include an optional absorption term to model frequency-dependent attenuation. k-Wave employs a pseudospectral method, which uses the Fast Fourier Transform (FFT) to compute spatial derivatives.

### 2.3.2 Hybrid k-Wave and Angular Spectrum Method (kW-ASM)

To improve computational efficiency, a hybrid method is used, combining k-Wave for the initial simulation of acoustic fields from the curved transducer to a planar surface near the skull and ASM for further propagation through the skull into the target brain region (Figure 3C). k-Wave is employed to simulate the acoustic field generated by the curved transducer up to a planar surface near the skull. The acoustic pressure field at this plane is defined as:

$$p_{k-Wave}(x, y, z = 0) = A \cdot e^{j\varphi}, \tag{13}$$

where $A$ and $\varphi$ represented as the amplitude of acoustic field at the boundary respectively as

$$A = |p_{k-Wave}(x, y, z = 0)|, \tag{14}$$

$$\varphi = \arg(p_{k-Wave}(x, y, z = 0)). \tag{15}$$

Once the pressure field at the planar surface is obtained, ASM is used to propagate the wave through the skull to the target region. The pressure field in the spatial frequency domain is expressed as

$$P(k_x, k_y) = \mathcal{F}\{p_{curve}(x, y, z = 0)\}, \tag{16}$$

where $\mathcal{F}$ represents the Fourier transform. The propagation in the z-direction is given by

$$P(x, y, z) = \iint P(k_x, k_y) e^{j\sqrt{k^2 - k_x^2 - k_y^2}\, z} \, dk_x dk_y, \tag{17}$$

where $k = 2\pi/\lambda$ is the wavenumber in the medium, the term $\sqrt{k^2 - k_x^2 - k_y^2}$ accounts for the phase change during propagation. Finally, the pressure field in the spatial domain at a distance $z_f$ from the transducer is obtained by applying the inverse Fourier transform

$$p(x, y, z_f) = \mathcal{F}^{-1}\{P(x, y, z_f)\}. \tag{18}$$

This hybrid approach reduces computational costs by limiting k-Wave simulations to the region near the transducer, while ASM efficiently handles propagation through the skull.

### 2.3.3 Hybrid Rayleigh-Sommerfeld Diffraction Integral and Angular Spectrum Method (RS-ASM)

This approach substitutes the k-Wave simulation with the Rayleigh-Sommerfeld (RS) Diffraction Integral for the initial field calculation, combining it with ASM for propagation through the skull. The RS diffraction integral models the acoustic field from the curved transducer to the planar surface near the skull. The pressure field is computed as

$$p(x, y, z) = \frac{1}{2\pi} \iint \frac{p_0}{r} \left(1 - \frac{jkr}{1+jkr}\right) e^{ikr}, \quad (19)$$

where $p_0$ is the pressure at the source point, $r = \sqrt{(x - x_0)^2 + (y - y_0)^2 + z^2}$ is the distance from source to target. After the planar pressure field is calculated using RS, ASM is applied to propagate the acoustic field through the skull to the target region, following the same steps as described in Section 2.3.2. This approach avoids the computational burden of k-Wave by using RS for the initial field calculation, providing an efficient and accurate alternative for modeling the curved transducer's field.

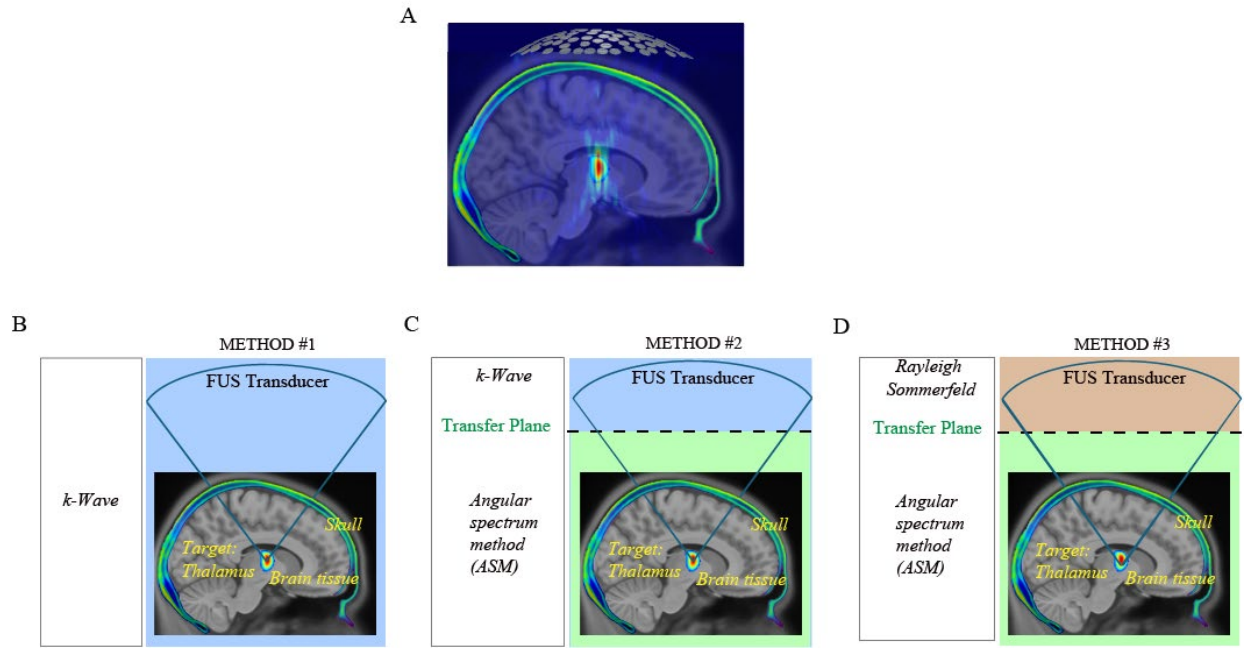

**Figure 3. The framework of the transcranial focused ultrasound simulation, targeting the thalamus as the deep brain focus.** (A) Displays the targeting setup, with the thalamus highlighted and the blue-green region representing the skull's sound speed map. (B) Shows the simulation using the standard k-Wave method
10

(Method #1). (C) Depicts the first proposed hybrid method (kW-ASM), combining k-Wave with the ASM (Method #2). (D) Illustrates the second proposed hybrid method (RS-ASM), integrating the Rayleigh-Sommerfeld diffraction integral with the ASM (Method #3). Each method is represented with its approach to modeling the acoustic field through the skull and brain tissue.

Figure 3 presents the framework of the three tFUS simulation methods applied in this study. Considering the advantage of the transcranial focusing in reaching deep brain regions, we selected the thalamus in the deep middle area of the brain as the target. The specific location of the thalamus was determined by the patient's MR image, and the skull acoustic properties were acquired from the rCT or sCT image. We extracted the skull from the CT image and mapped the speed image, followed by combining the skull with MR image as the schematic diagram (Figure 3A, 3B, 3C and 3D).

### 2.3.4 Statistics Analysis

Linear regression analysis was performed on these paired values using GraphPad Prism (Version 10.1.2), alongside a correlation analysis, to evaluate the degree of agreement between the two datasets. The coefficient of determination Square of R ($R^2$) was calculated. Statistical significance was evaluated by the unpaired two-tailed student t-test using GraphPad Prism (Version 10.1.2), and p-value < 0.05 was defined as statistically significant.

## 2. Results

### 3.1 Validation of sCT Accuracy Against Real CT for tFUS Planning

To establish the feasibility of MRI-derived sCT as a substitute for rCT in tFUS planning, we conducted a comprehensive evaluation of its accuracy across five skull samples (Skull I–V). The assessment included direct structural comparison, voxel-wise intensity correlation, and quantitative error analysis.

Figure 4 illustrates the structural comparison between MRI, rCT, and sCT images. The 3D reconstructions from MRI (Figure 4A) depict the anatomical variability among the five skulls, while axial slices of rCT and sCT (Figure 4B) reveal high structural fidelity between the synthetic and real CT images, particularly in regions of the skull relevant to acoustic modeling. Absolute difference maps (Figure 4C) indicate that the discrepancies between sCT and rCT are primarily localized in soft tissue regions, whereas the bone structures exhibit minimal differences. To further quantify the fidelity of sCT to rCT, voxel intensity distributions were analyzed. Figure 5A shows a linear regression analysis between rCT and sCT intensities, yielding a high correlation coefficient ($R^2 = 0.9253$, $P < 0.0001$), demonstrating that the sCT preserves key Hounsfield Unit (HU) distributions. Figure 5B presents edge-enhanced projections of rCT and sCT, further illustrating the structural consistency across different skull samples.

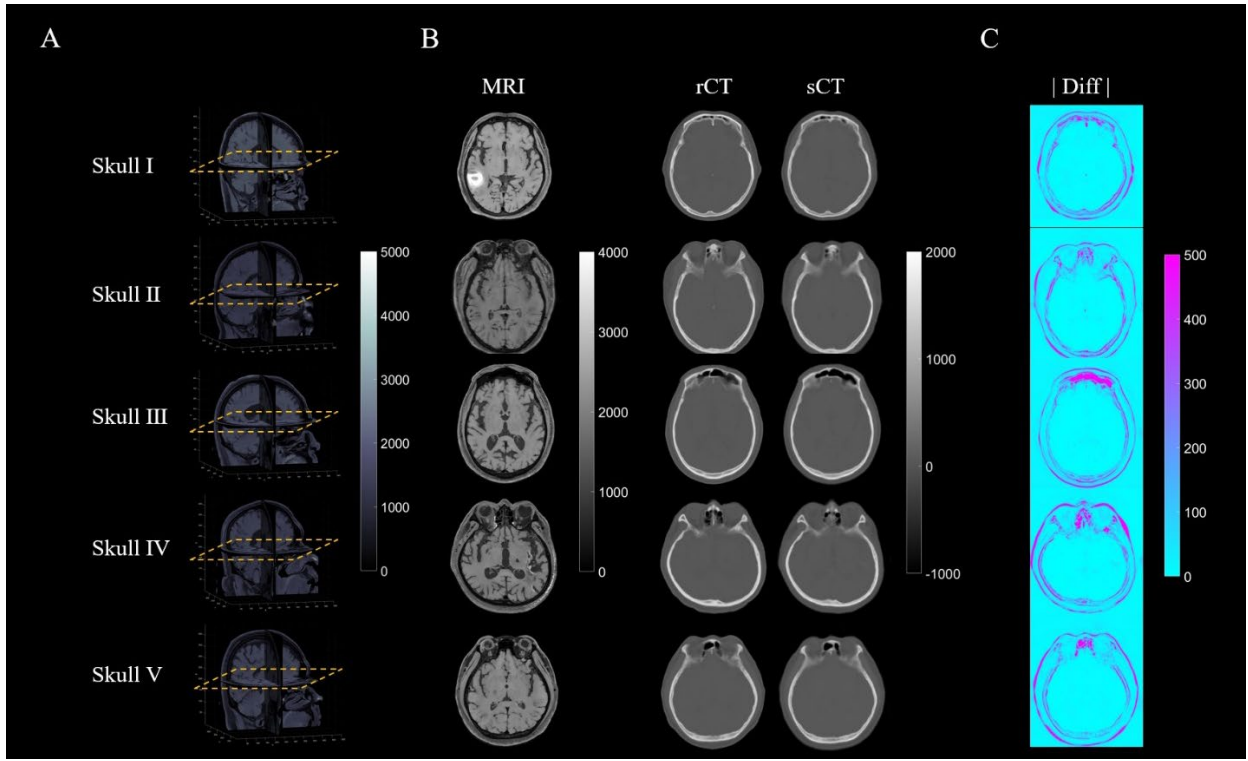

**Figure 4. Comparison of MRI, rCT and sCT.** (A) shows the 3D reconstruction of MRI data for five different skulls (Skull I to Skull V). (B) compares axial slices of MRI, real CT, and sCT images, highlighting the structural similarities between sCT and real CT generated by the proposed algorithm. (C) depicts the absolute difference maps (|Diff|) between the real CT and sCT images, quantifying the deviations in Hounsfield Units across the skull regions. This comparison demonstrates the accuracy of the sCT reconstruction in replicating key acoustic properties of the skull.

The mean absolute error (MAE) was computed to assess the deviations between rCT and sCT in different anatomical regions. Table 1 summarizes the results, comparing the proposed method to a conventional pseudo-CT approach. Across all metrics, the proposed sCT generation method demonstrates improved accuracy. The proposed MRI-derived sCT demonstrated improved accuracy compared to the conventional pseudo-CT approach, with a lower mean absolute error (MAE) across different anatomical regions. Specifically, the overall head MAE was reduced from 158.42 HU to 99.85 HU, while the brain region ($0 \leq HU \leq 100$) showed a decrease from 64.93 HU to 44.38 HU, indicating better soft tissue estimation. In the skull region ($HU > 2000$), the MAE was significantly reduced from 317.13 HU to 177.71 HU, highlighting improved bone density reconstruction. These findings suggest that sCT effectively preserves key acoustic properties required for transcranial ultrasound simulations while eliminating the need for ionizing radiation.

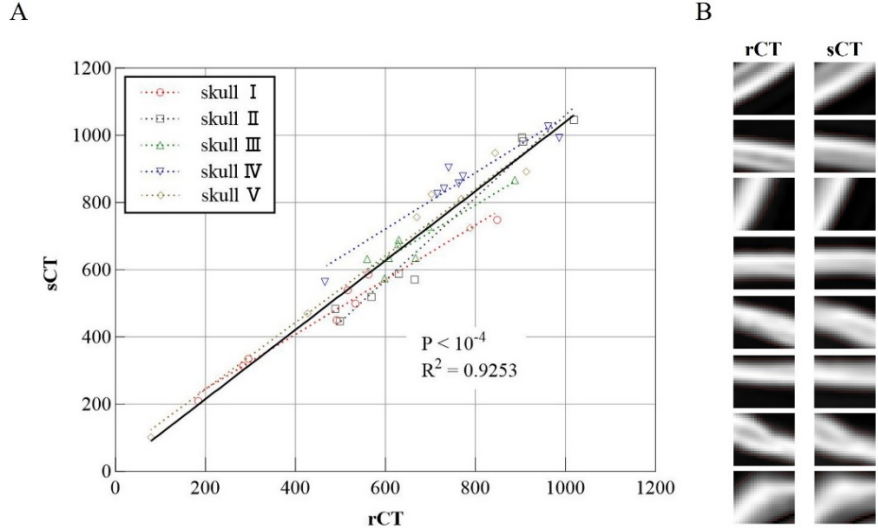

**Figure 5. Comparison of rCT and MRI-Derived sCT in skull analysis.** (A) Presents a correlation plot comparing the mean density (Hounsfield Units) between rCT and sCT across five skulls (Skull I to Skull V), showing a strong linear relationship ($R^2 = 0.9253, P < 0.0001$). (B) Highlights eight selected regions of interest (ROIs) from one representative skull, comparing structural details between rCT and sCT images, demonstrating the consistency of sCT in capturing key skull features.

These results indicate that MRI-derived sCT provides a viable alternative to rCT, with reduced MAE across skull and brain regions. The strong correlation in HU values and the low structural deviation confirm the applicability of sCT for estimating acoustic properties relevant to tFUS. The ability to accurately reconstruct skull density and speed of sound from MRI eliminates the need for CT-based imaging, enabling a radiation-free workflow for tFUS planning. In the following sections, we further evaluate the integration of sCT with rapid simulation methods to assess its impact on computational efficiency and focal targeting accuracy.

Table 1: Synthesized CT evaluation on the test set (N=5).

| Method | Head MAE↓ | Brain MAE↓ ($0 \leq HU \leq 100$) | Skull MAE↓ ($0 \leq HU \leq 2000$) |
|---|---|---|---|
| Pseudo-CT [54] | 158.42 | 64.93 | 317.13 |
| This study | 99.85 | 44.38 | 177.71 |

**3.2 Validation of sCT-Based Simulations for Transcranial Focused Ultrasound Planning**

Building upon the anatomical and acoustic consistency of sCT demonstrated in Section 3.1, we further evaluated its effectiveness in focused ultrasound simulations. This section investigates whether sCT can (1) replicate the acoustic propagation properties derived from rCT using the k-Wave solver, (2) be reliably integrated into rapid simulation methods such as kW-ASM and RS-ASM, and (3) achieve accurate targeting and efficient computation in quantitative terms. In section 3.2, we compared six representative simulation pipelines: (1) k-Wave with real CT (kWave-rCT), (2) k-Wave with sCT (kWave-sCT), (3) k-Wave combined with Angular Spectrum Method using sCT (kW-ASM-sCT), (4) Rayleigh-Sommerfeld Angular Spectrum Method using sCT (RS-ASM-sCT), (5) k-Wave combined with Angular Spectrum Method using rCT (kW-ASM-rCT), and (6) RS-ASM using real CT (RS-ASM-rCT). Results are summarized in Figures 6–8.

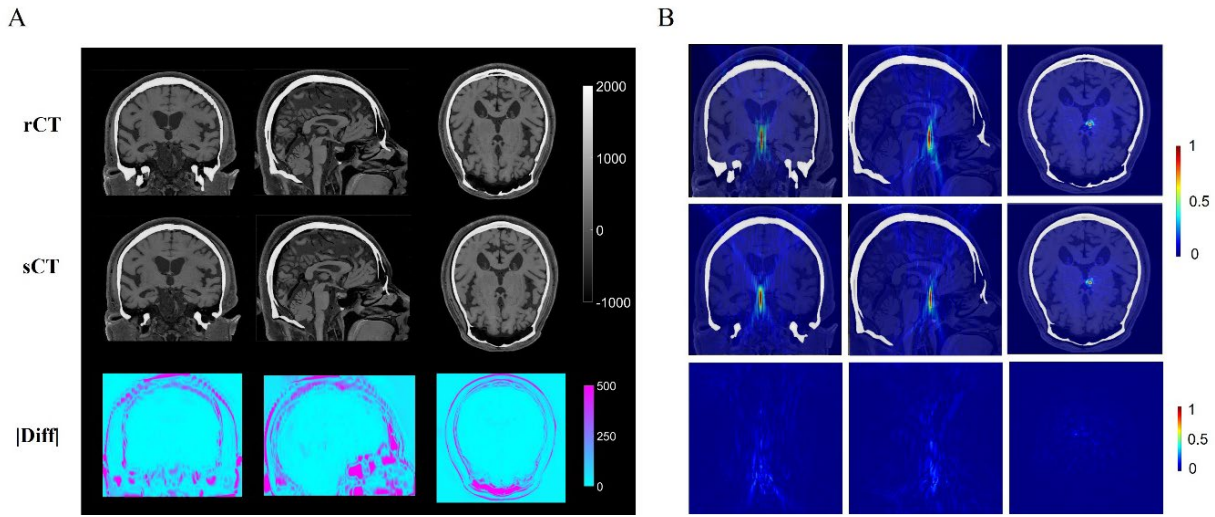

**Figure 6. Comparison of rCT and sCT in transcranial Ultrasound Simulations Using k-Wave.** (A) Displays rCT and sCT images overlaid with MRI in coronal, sagittal, and horizontal planes, along with absolute difference maps (|Diff|), showing minimal deviations in soft tissue and high consistency in bone structures. (B) Presents transcranial ultrasound focusing simulations using the k-Wave method, showing near-identical acoustic field distributions for rCT and sCT data across all planes, validating sCT's accuracy for transcranial ultrasound simulations.

### 3.2.1 Acoustic Field Agreement between sCT and rCT

To assess the acoustic modeling fidelity of sCT, we performed full-wave simulations using the k-Wave toolbox on both rCT- and sCT-derived skull models under identical conditions. The simulation targeted deep brain structures near the hypothalamus, and key parameters such as transducer aperture, focus depth, and spatial resolution were matched across cases. Figure 6A displays multi-view structural comparisons in the coronal, sagittal, and axial planes. Both rCT and sCT exhibit clear delineation of cranial bone and intracranial anatomy. The overlaid MRI confirms anatomical correspondence, while the absolute HU difference maps highlight that intensity deviations were primarily localized at the skull base and soft tissue boundaries. In contrast, the cranial vault—most relevant for transcranial acoustic transmission—exhibited high agreement, typically with <100 HU error. Corresponding k-Wave simulations (Figure 6B) showed that sCT-derived pressure fields closely replicated those from rCT. The normalized pressure maps reveal overlapping beam trajectories and well-focused spots in all planes. Absolute pressure difference maps, mostly below 0.2 in normalized units, showed minor deviations concentrated in peripheral regions, with negligible differences in the focal zone. These results support that sCT can serve as a structurally and acoustically accurate input for full-wave tFUS modeling.

### 3.2.2 Validation of sCT with Rapid Simulation Methods

To determine whether sCT is compatible with rapid modeling techniques, we simulated deep and shallow brain targets using sCT as the propagation medium with two accelerated methods: kW-ASM and RS-ASM. Each was benchmarked against k-Wave simulations using rCT, which served as the gold standard. In this study, both deep and shallow transcranial focusing were performed. For deep focusing, a sensor array composed of 128 elements (arranged in a 2×2 configuration) was utilized. The position of each element was determined based on localization data obtained under actual wearing conditions. The focal depth was set at 8 cm beneath the cranial vault, corresponding to the thalamus. For shallow focusing, a single-element bowl-shaped transducer was employed, with a radius of $6 \times 10^{-2} m$ and a diameter of $8.84 \times 10^{-2} m$. The focal depth was set at 4 cm beneath the scalp. Ultrasound frequency was 500kHz for both shallow- and deep-focus settings.

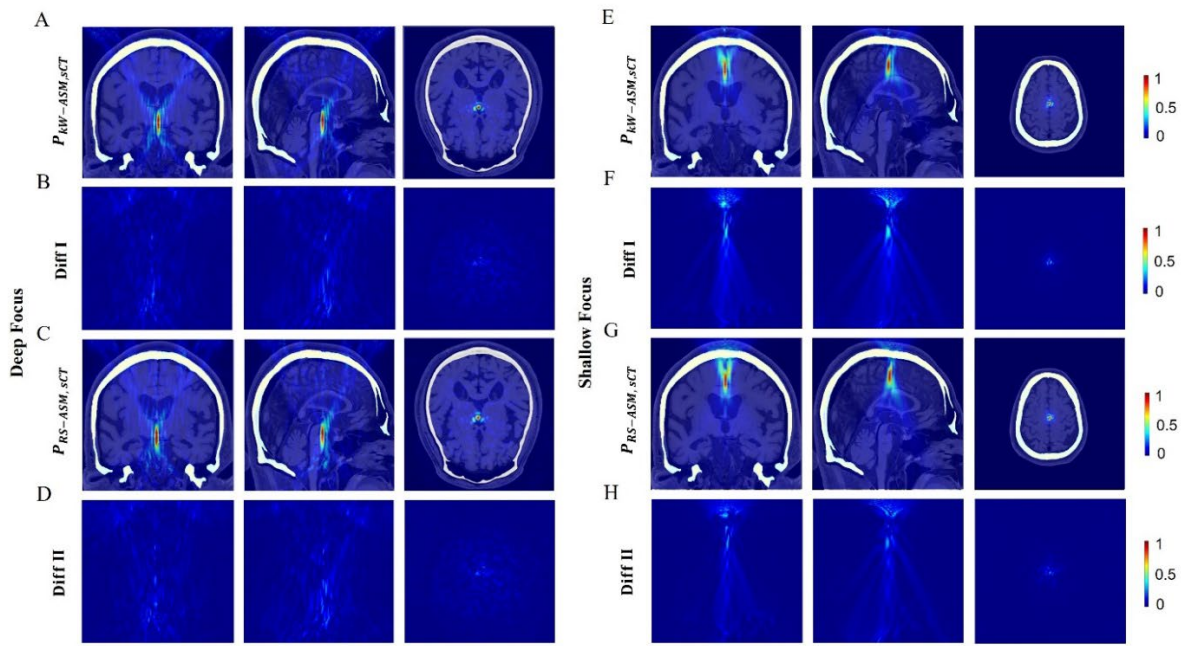

**Figure 7. Simulated pressure field distributions using kW-ASM and RS-ASM methods with synthetic CT (sCT) for deep and shallow transcranial ultrasound (tFUS) focusing.** (A, C) Show the normalized pressure distributions obtained by kW-ASM-sCT and RS-ASM-sCT, respectively, for the deep-focus case. (E, G) Present the corresponding results for the shallow-focus case. (B, D, F, H) Illustrate the absolute difference maps (Diff I and Diff II) between the respective methods and the k-Wave gold standard. The colormap represents normalized acoustic pressure (0–1) for the pressure fields and 0–1 for the difference maps.

For deep-focus targeting, kW-ASM-sCT generated pressure distributions with high consistency relative to kWave-rCT (Figure 7A), including tight focal spots and minimal side lobes. The difference maps (Figure 7B) revealed small peripheral deviations, with the focus remaining virtually unaffected. RS-ASM-sCT also maintained good spatial agreement (Figure 7C), though the associated error map (Figure 7D) showed slightly more diffused off-axis differences. For shallow-focus simulations, both rapid methods continued to demonstrate high fidelity. As shown in Figure 7E, kW-ASM-sCT produced coherent beam convergence with a well-defined superficial focus. The corresponding error (Figure 7F) was low and mostly located at the transducer-facing skull surface. RS-ASM-sCT also reproduced the intended focal region (Figure 7G), with Figure 7H showing minor lateral discrepancies likely due to wavefront sensitivity in this method. Together, these results confirm that sCT, when combined with rapid solvers, accurately preserves the beam's trajectory and intensity profile across a range of targeting depths. This supports its potential use in fast, CT-free simulation workflows for real-time treatment planning.

### 3.2.3 Quantitative Comparison of Focal Accuracy and Simulation Speed

To quantify the performance of sCT-based simulations in terms of targeting accuracy and computational efficiency, we evaluated six simulation pipelines by conducting both deep and shallow focusing across five different skull models, plus an additional skull-free configuration as reference. The six pipelines included: (1) kWave-rCT (reference full-wave simulation with real CT), (2) kWave-sCT (full-wave with synthetic CT), (3) kW-ASM-CT, (4) kW-ASM-sCT, (5) RS-ASM-CT, and (6) RS-ASM-sCT. Each pipeline was tested under two focusing depths (shallow and deep) across six anatomical settings (five individual skulls and one no-skull condition), resulting in 12 simulation scenarios per pipeline. This design allowed us to isolate the effect of skull presence and anatomical variability on beam characteristics. For each scenario, we recorded five evaluation metrics: axial focal length (FLHM), longitudinal error in the sagittal plane, lateral focal width (FWHM), transverse error in the axial plane, and total simulation time.

As shown in Figure 8A and 8B, the comparison of axial focal length and longitudinal deviation revealed that all pipelines produced similar FLHM values, with kWave-rCT yielding $20 \pm 4.1$ mm. Other pipelines produced consistent focal lengths around 19 mm, with standard deviations ranging from ±4.6 to ±5.3 mm, indicating that axial beam shaping was preserved regardless of the solver or input image type. In terms of longitudinal deviation, kW-ASM-CT achieved the highest axial accuracy (1.5 ± 1.2 mm), followed by kW-ASM-sCT (2.2 ± 1.3 mm), RS-ASM-CT (2.2 ± 1.5 mm), RS-ASM-sCT (2.6 ± 1.6 mm), and kWave-sCT (2.5 ± 1.5 mm). These results highlight that ASM-based pipelines better maintain axial focal position, particularly when combined with sCT.

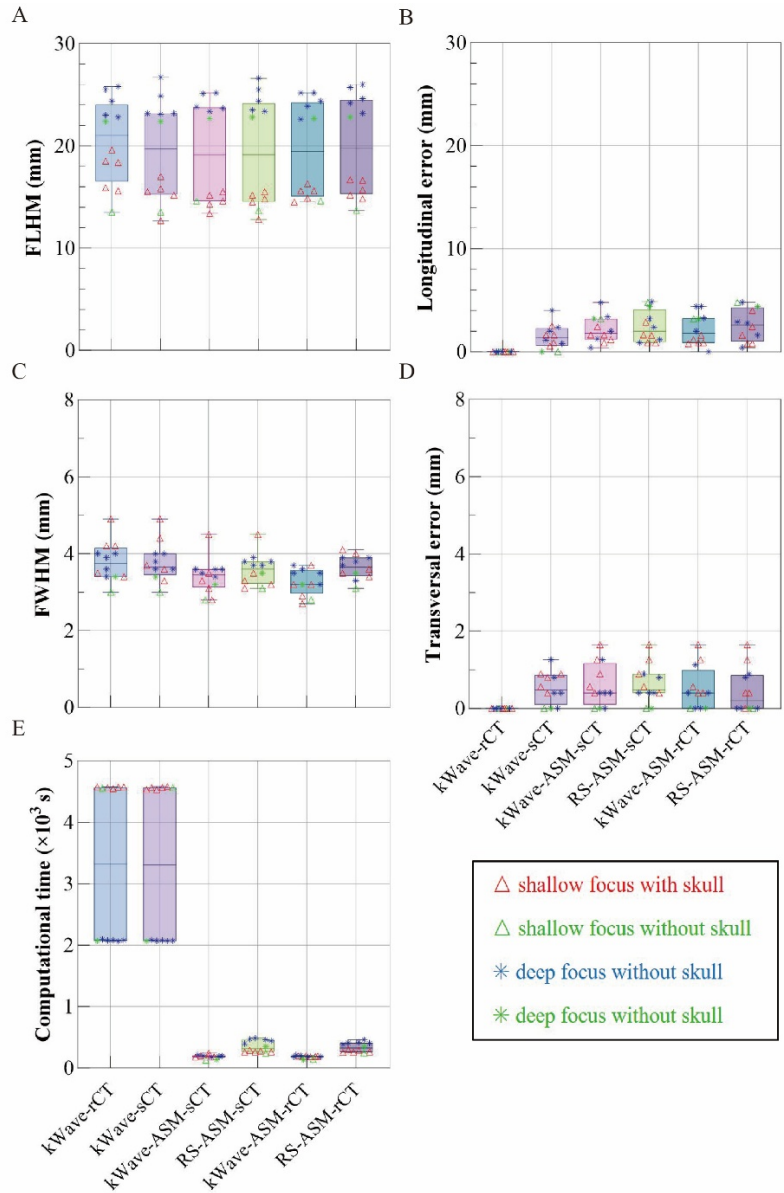

**Figure 8. Quantitative comparison of six transcranial focused ultrasound (tFUS) simulation pipelines across five skull models, using kWave with real CT (kWave-rCT) as the reference.** Simulations were conducted for deep and shallow targets, with and without skulls. (A) Axial focal length (FLHM): kWave-

rCT yielded 20 ± 4.1 mm; other pipelines produced similar results: 19 ± 4.9 mm (kWave-sCT), 19 ± 5.3 mm (kW-ASM-CT), 19 ± 4.6 mm (kW-ASM-sCT), 19 ± 4.8 mm (RS-ASM-CT), 19 ± 4.9 mm (RS-ASM-sCT). (B) Longitudinal error (sagittal plane): Mean deviations from kWave-rCT were: 1.5 ± 1.2 mm (kW-ASM-CT), 2.2 ± 1.3 mm (kW-ASM-sCT), 2.5 ± 1.5 mm (kWave-sCT), 2.2 ± 1.5 mm (RS-ASM-CT), 2.6 ± 1.6 mm (RS-ASM-sCT). (C) Lateral focal width (FWHM): 3.8 ± 0.51 mm (kWave-rCT), 3.8 ± 0.51 mm (kWave-sCT), 3.4 ± 0.45 mm (kW-ASM-CT), 3.6 ± 0.40 mm (kW-ASM-sCT), 3.3 ± 0.34 mm (RS-ASM-CT), 3.7 ± 0.30 mm (RS-ASM-sCT). (D) Transverse error (horizontal plane): Lateral deviations from reference were: 0.54 ± 0.41 mm (kWave-sCT), 0.60 ± 0.55 mm (kW-ASM-sCT), 0.64 ± 0.49 mm (RS-ASM-sCT), 0.52 ± 0.55 mm (kW-ASM-CT), 0.45 ± 0.58 mm (RS-ASM-CT). (E) Simulation time: Full-wave simulations (kWave-rCT/sCT) required ~3320 s, while kW-ASM-rCT/sCT and RS-ASM-rCT/sCT completed in 187 ± 27 s and 345 ± 85 s, respectively, achieving ~94% time savings for kW-ASM and ~90% for RS-ASM.

As shown in Figure 8C and 8D, the lateral focal width (FWHM) and transverse deviation were also consistent across pipelines. The FWHM values ranged from 3.3 to 3.8 mm. Both kWave-rCT and kWave-sCT showed 3.8 ± 0.51 mm, while kW-ASM-CT (3.4 ± 0.45 mm) and RS-ASM-CT (3.3 ± 0.34 mm) produced slightly narrower beams. sCT-based ASM pipelines also maintained good lateral precision: 3.6 ± 0.40 mm for kW-ASM-sCT and 3.7 ± 0.30 mm for RS-ASM-sCT. Regarding transverse positioning (horizontal plane), all methods achieved sub-millimeter accuracy. RS-ASM-CT had the lowest mean deviation (0.45 ± 0.58 mm), followed closely by kW-ASM-CT (0.52 ± 0.55 mm), kWave-sCT (0.54 ± 0.41 mm), kW-ASM-sCT (0.60 ± 0.55 mm), and RS-ASM-sCT (0.64 ± 0.49 mm). These findings confirm the geometric fidelity of the spectral-domain approaches.

As shown in Figure 8E, the computational performance differed markedly. Full-wave simulations using kWave with CT or sCT required approximately 3320 seconds per run. In contrast, kW-ASM pipelines completed in 187 ± 27 seconds, and RS-ASM in 345 ± 85 seconds, corresponding to ~94% and ~90% reductions in runtime, respectively. These findings confirm that MRI-derived sCT, when combined with kW-ASM or RS-ASM solvers, provides a highly accurate and computationally efficient strategy for tFUS simulation. The framework not only eliminates radiation exposure but also achieves reliable spatial accuracy, supporting its translational potential for clinical neuromodulation planning.

## 4 Discussion

This study presents a novel simulation framework that integrates MRI-derived synthetic CT with rapid ultrasound modeling techniques to support accurate and efficient transcranial focused ultrasound treatment planning. Our findings demonstrate that sCT can accurately approximate real CT in terms of both anatomical fidelity and acoustic property estimation, and that sCT is compatible with both full-wave and accelerated simulation solvers.

The structural fidelity of sCT compared to rCT was comprehensively validated through voxel-wise correlation ($R^2$ = 0.9253, $P < 0.0001$), regional Hounsfield Unit (HU) comparisons, and MAE analysis. Notably, the skull-specific MAE decreased to 177.71 HU with the proposed model, a substantial improvement over previous pseudo-CT approach as illustrated in [86]. The bone structure, which governs ultrasound transmission and phase aberration, was accurately preserved across multiple skull samples, as evidenced by intensity maps and ROI visualizations. These results confirm that sCT is sufficiently accurate for estimating density and speed of sound distributions in skull models, offering a radiation-free alternative for tFUS simulation pipelines. Using the k-Wave full-wave solver, we demonstrated that sCT-based simulations closely replicate those using rCT in terms of pressure field distribution, focal location, and beam morphology. For deep brain targets near the hypothalamus, the normalized pressure fields derived from sCT and rCT were nearly identical across coronal, sagittal, and axial planes. The absolute error in pressure (|Diff|) remained below 0.2 (normalized units) and was confined to peripheral regions, with negligible discrepancy in the focal zone (Figure 6). This high degree of concordance reinforces the suitability of sCT as a substitute for rCT in full-wave acoustic modeling, especially in scenarios where repeated exposure to ionizing radiation is not desirable.

Beyond full-wave modeling, we further assessed the utility of sCT in combination with two rapid simulation methods: kW-ASM and RS-ASM. These approaches substantially reduce computational time by decomposing the simulation pipeline into transducer-field and skull-propagation stages. Results showed that for both deep and shallow focus cases, the beam patterns generated using sCT matched those from the kWave-CT gold standard in both spatial distribution and peak intensity. Difference maps (Figure 7) revealed that the majority of discrepancies occurred outside the focal region and were minor in magnitude, indicating that neither skull heterogeneity nor target depth substantially impaired accuracy. Importantly, RS-ASM exhibited slightly larger variance in lateral regions compared to kW-ASM, which may be attributed to its heightened sensitivity to wavefront curvature during propagation. Nevertheless, both solvers preserved focal integrity and proved robust against minor deviations in input skull data. This finding confirms that sCT can be flexibly incorporated into rapid modeling methods, providing the foundation for scalable and real-time tFUS treatment planning.

Quantitative analysis across five metrics (Figure 8) reinforced the suitability of sCT-based pipelines. As shown in Figure 8A–B, axial focal length and sagittal deviation remained consistent across all methods, with kW-ASM-sCT and RS-ASM-sCT achieving sub-2.6 mm axial deviations. As shown in Figure 8C–D, lateral beam widths were comparable (~3.3–3.8 mm), and transverse localization errors stayed below 0.7 mm. Most notably, as shown in Figure 8E, the kW-ASM and RS-ASM methods reduced simulation time from ~3320 s to 187 ± 27 s and 345 ± 85 s, respectively corresponding to ~94% and ~90% savings. These computational advantages are especially critical for real-time or iterative clinical applications, including focal adjustment, phase correction, or multi-target navigation.

Despite these promising results, several limitations should be acknowledged. First, although sCT closely approximates rCT, residual HU errors in regions with complex anatomical structures (e.g., skull base) may still affect acoustic simulation in a small subset of trajectories. Second, rapid simulation methods inherently involve approximations, such as planar propagation assumptions and frequency domain simplifications, which could reduce modeling accuracy under extreme geometrical conditions. Future work should focus on three key directions: (1) enhancing the generalizability of the sCT model across diverse populations and imaging protocols, (2) incorporating skull anisotropy and attenuation models directly into the sCT-to-acoustics pipeline, and (3) validating the framework on in vivo or clinical tFUS datasets with functional or behavioral endpoints. Additionally, integrating this framework with treatment navigation systems could enable end-to-end CT-free ultrasound planning.

## 5  Conclusion

We have developed and validated a rapid, CT-free simulation framework for transcranial focused ultrasound targeting by integrating MRI-derived synthetic CT with both full-wave and accelerated modeling methods. The proposed approach demonstrated strong agreement with CT-based simulations in terms of structural fidelity, focal accuracy, and beam morphology, while offering substantial improvements in computational efficiency. The kW-ASM and RS-ASM solvers, when combined with sCT, achieved >90% reductions in simulation time with sub-millimeter targeting accuracy. These results highlight the feasibility of sCT-based planning for real-time and scalable clinical tFUS applications. This framework has the potential to facilitate individualized neuromodulation and therapeutic ultrasound interventions without the need for radiation exposure, and to accelerate the translation of ultrasound-based brain stimulation into routine clinical practice.